# The magnetic orders induced ferroelectricity

C. D. Hu

Department of Physics, National Taiwan University, Taipei, Taiwan R. O. C.

Abstract

We studied the ferroelectricity of magnetic oxides in which its emergence coincides with the onset of a second incommensurate magnetic order. We solved for the wave function of $e_g$ electrons in the presence of magnetic orders. The coupling between the magnetic and electric orders is provided by the spin-orbit interaction. It was found that the net electric dipole moment of the system came from the bond between the transition metal and oxygen atoms. The anisotropy of d-orbitals also played an important role. Finally, the form of the coupling leads us to the conclusion that it is an improper ferroelectricity.

The multiferroic phenomena had been studied for a long time[1]. The interest has been revived by recent experimental findings. They showed that magnetic and ferroelectric orders are closely related[2-5]. For example, TbMnO$_3$ undergoes antiferromagnetic phase transition at 41K at which an incommensurate sinusoidal magnetic order begins to develop. Below 28K which was often referred as $T_{lock}$, a second magnetic order was observed[6]. This occurrence coincides with the divergence of dielectric constant and the emerging of electric dipole moment of the system. It is this fascinating interplay between ferroelectric and magnetic orders that has attracted many researchers. As far as we know, the mechanism of the coupling between the incommensurate magnetic orders and ferroelectricity is still an unsolved topic. There have been calculations of electric polarization based on atomic displacements or lattice distortions[7,8]. There are also proposals of dipole moment due to electronic distribution[9,10]. However, it is interesting to note that the spin-orbit interaction played a crucial role in both schools. Sergienko and Dagotto[8] espoused the idea that ferroelectricity is produced by Dzyaloshinskii-Moriya(DM) interaction[11,12]. Katsura et. al.[10] showed that magnetoelectric effect was induced by spin current via spin-orbit coupling. At this stage, there is no well-developed microscopic theory of multiferroic phenomena. In refs. 8 and 9 phase diagrams were drawn and the regions of ferroelectricity were identified. However, it is still not clear that how magnetic orders are coupled to ferroelectricity. If it is due to DM interaction, then the question remains as what the driving force is. Is it from the displacement of atoms as it was suggested in ref. 8? And if so, why is there linear relation between the displacement and the coupling constant $D$? Or does it come from an "internal electric field" as it was proposed in ref. 10? A theory of spin current coupling to the motion of charge is presented. The computation is based on only two noncollinear spins. Then, the question rises as how the mechanism works in a collective manner, i. e. in a crystal.

In this work we tried to give a global view. We proposed a possible mechanism of coupling between the magnetic orders and the ferroelectric order. The systems we had in mind were orthorhombic perovskites such as TbMnO$_3$. These are complex systems in which Hund's coupling, anisotropy of orbitals and lattice distortion all have significant effect. The spin-orbit interaction also plays a subtle role. Thus, our Hamiltonian has the form $H = H_0 + H_1$ with the latter been the spin-orbit interaction and,

$$H_0 = \sum \varepsilon_p c^\dagger_{i,lp\sigma} c_{i,lp\sigma} + \sum \varepsilon_d c^\dagger_{j,d\sigma} c_{j,d\sigma} + \sum t_{l,ij}(c^\dagger_{i,lp\sigma} c_{j,d\sigma} + c^\dagger_{j,d\sigma} c_{i,lp\sigma}) - J \sum S_j \bullet s_j \qquad 1$$

Here $\varepsilon_p$, $\varepsilon_d$, $c_{i,lp\sigma}$ and $c_{j,d\sigma}$ are the energies and field operators of the electrons in the p-orbitals in oxygen and d-orbitals in transition element ions with $\sigma$ being the spin index and $l$ the oxygen atom index in the basis. To simplify the computation we considered the dynamics on the *ab* plane. We also made the approximation that the lattice constants *a* and *b* are equal. As a result, there are one transition element atom and two oxygen atoms in our basis. Let one M(metal)-O ($l = x$) bond be along x-direction and the other ($l = y$) along y-direction. Thus the oxygen atoms are at $R_j + r_x$ and $R_j + r_y$ in the j-th unit cell where $r_x = \hat{x}b/2\sqrt{2}$ and $r_y = \hat{y}b/2\sqrt{2}$ with $b/\sqrt{2}$ being our reduced lattice constant. The orbitals we chose were the most active ones. For example, the $p_x$ orbitals of oxygen atoms were incorporated if the M-O bond was along x-direction. and $p_y$ orbitals for y-direction bonds. For the same reason we considered only one e$_g$ orbital. There are two e$_g$ orbitals. Due to the Jahn-Teller lattice distortion the degeneracy is lifted. The lower state (as well as the upper) has the form $A|x^2 - y^2\rangle + B|3z^2 - r^2\rangle$ and this is the state we considered. We later found out that the values of $A$ and $B$ were not going to affect qualitatively our results. The tight-binding matrix elements $t_{l,ij}$ can be found in Slater and Koster[13]: $t_{l,ij} = (-1)^{i-j}(\Lambda_l A + B/\sqrt{3})(pd\sigma)$ where $\Lambda_l = -1(1)$ if $l = x(y)$. The fourth term is the Hund's coupling. $S_j$ and $s_j$ denote the spins of the t$_{2g}$ and e$_g$ electrons. The t$_{2g}$ electrons of metal ions are relatively inert since their energy is several electron volts lower than that of e$_g$ electrons. They form a $S = 3/2$ state. Their effect is to provide a fictitious magnetic field to align the spins of e$_g$ electrons.

From the start, we assume that there is a spiral magnetic order due to the t$_{2g}$ electrons. Neutron scattering data[6] indicate that this is indeed this case below $T_{lock}$. The spins of t$_{2g}$ electrons can be expressed as

$$S_j^x = S_0^x \cos(\mathbf{q} \cdot \mathbf{R}_j + \gamma)$$
$$S_j^y = S_0^y \cos(\mathbf{q} \cdot \mathbf{R}_j + \gamma) \qquad 2$$
$$S_j^z = S_0^z \cos(\mathbf{q} \cdot \mathbf{R}_j)$$

where $\mathbf{q}$ is the wave vector of the incommensurate order and $\gamma$ is the phase difference between the oscillations of $S^z$ and $S^{x(y)}$.

We can then take the Fourier transform of the Hamiltonian in eq. (1) and reach

$$\begin{aligned}H_0 =& \sum \varepsilon_p c_{\mathbf{k},lp\sigma}^\dagger c_{\mathbf{k},lp\sigma} + \sum \varepsilon_d c_{\mathbf{k},d\sigma}^\dagger c_{\mathbf{k},d\sigma} + \sum (V_\mathbf{k}^l c_{\mathbf{k},lp\sigma}^\dagger c_{\mathbf{k},d\sigma} + V_\mathbf{k}^{l*} c_{\mathbf{k},d\sigma}^\dagger c_{\mathbf{k},lp\sigma}) \\ & - (J/2) \sum [S_0^z (c_{\mathbf{k},d\uparrow}^\dagger c_{\mathbf{k+q},d\uparrow} - c_{\mathbf{k},d\downarrow}^\dagger c_{\mathbf{k+q},d\downarrow}) + S_0^+ e^{i\gamma} c_{\mathbf{k},d\downarrow}^\dagger c_{\mathbf{k+q},d\uparrow} + S_0^- e^{i\gamma} c_{\mathbf{k},d\uparrow}^\dagger c_{\mathbf{k+q},d\downarrow} + H.c.]\end{aligned} \qquad 3$$

where $S_0^\pm = S_0^x \pm i S_0^y$ and $V_\mathbf{k}^l = (\Lambda_l A + B/\sqrt{3})(1 - e^{i\mathbf{k}\cdot\mathbf{r}_l})(pd\sigma)$ with $l = x(y)$. Thus, we have a Hamiltonian similar to that of spin density wave. There are twelve coupled states, denoted by $c_{\mathbf{k},d\sigma}$, $c_{\mathbf{k+q},d\sigma}$, $c_{\mathbf{k},lp\sigma}$ and $c_{\mathbf{k+q},lp\sigma}$. With these twelve states as basis, we wrote the Hamiltonian as a 12×12 matrix:

$$\begin{pmatrix} \varepsilon_d & 0 & -JS_{z0}/2 & -J'^* S_0^-/2 & V_\mathbf{k}^{x*} & 0 & V_\mathbf{k}^{y*} & 0 & 0 & 0 & 0' & 0 \\ 0 & \varepsilon_d & -J'^* S_0^+/2 & JS_{z0}/2 & 0 & V_\mathbf{k}^{x*} & 0 & V_\mathbf{k}^{y*} & 0 & 0 & 0 & 0 \\ -JS_{z0}/2 & -J' S_0^-/2 & \varepsilon_d & 0 & 0 & 0 & 0 & 0 & V_{\mathbf{k+q}}^{x*} & 0 & V_{\mathbf{k+q}}^{y*} & 0 \\ -J' S_0^+/2 & JS_{z0}/2 & 0 & \varepsilon_d & 0 & 0 & 0 & 0 & 0 & V_{\mathbf{k+q}}^{x*} & 0 & V_{\mathbf{k+q}}^{y*} \\ V_\mathbf{k}^x & 0 & 0 & 0 & \varepsilon_p & 0 & 0 & 0 & 0 & 0 & 0 & 0 \\ 0 & V_\mathbf{k}^x & 0 & 0 & 0 & \varepsilon_p & 0 & 0 & 0 & 0 & 0 & 0 \\ V_\mathbf{k}^y & 0 & 0 & 0 & 0 & 0 & \varepsilon_p & 0 & 0 & 0 & 0 & 0 \\ 0 & V_\mathbf{k}^y & 0 & 0 & 0 & 0 & 0 & \varepsilon_p & 0 & 0 & 0 & 0 \\ 0 & 0 & V_{\mathbf{k+q}}^x & 0 & 0 & 0 & 0 & 0 & \varepsilon_p & 0 & 0 & 0 \\ 0 & 0 & 0 & V_{\mathbf{k+q}}^x & 0 & 0 & 0 & 0 & 0 & \varepsilon_p & 0 & 0 \\ 0 & 0 & V_{\mathbf{k+q}}^y & 0 & 0 & 0 & 0 & 0 & 0 & 0 & \varepsilon_p & 0 \\ 0 & 0 & 0 & V_{\mathbf{k+q}}^y & 0 & 0 & 0 & 0 & 0 & 0 & 0 & \varepsilon_p \end{pmatrix} \qquad 4$$

where $J' = Je^{i\gamma}$.

The eigenvalues of the matrix, besides $E = \varepsilon_p$ with four-fold degeneracy, can be obtained by solving the secular equations

$$[(E - \varepsilon_p)(E - \varepsilon_d) - \sum_l |V_\mathbf{k}^l|^2][(E - \varepsilon_p)(E - \varepsilon_d) - \sum_l |V_{\mathbf{k+q}}^l|^2]$$
$$= [(E - \varepsilon_p)J/2]^2 (S_0^z \mp i\sqrt{S_0^+ S_0^-} e^{i\gamma})(S_0^z \pm i\sqrt{S_0^+ S_0^-} e^{-i\gamma}). \qquad 5$$

The roots can be found analytically. However, their expressions are rather lengthy. It is sufficient for our purpose to consider the approximate forms by using the fact that $(pd\sigma) \ll \varepsilon_d - J/2 - \varepsilon_p$. The unit cell in reciprocal space is halved due to the spin density wave-like interaction. Hence, the top two bands are empty if there are five valence electrons (two p-electrons for each oxygen atom and one $e_g$ electron) in a unit cell. The highest occupied state has the eigenvalue

$$E_\mathbf{k} \simeq \varepsilon_d - J|\Sigma|/2 + \frac{\sum_l (|V_\mathbf{k}^l|^2 + |V_{\mathbf{k+q}}^l|^2)}{2(\varepsilon_d - J|\Sigma|/2 - \varepsilon_p)} \qquad 6$$

where $\Sigma = S_0^z + i\sqrt{S_0^+ S_0^-}\, e^{i\gamma}$. The eigenstate is

$$|\Psi_{\mathbf{k}}\rangle = C \sum_\sigma [\alpha_\sigma c^\dagger_{\mathbf{k},d\sigma} + \beta_\sigma c^\dagger_{\mathbf{k+q},d\sigma} + \sum_l (V^l_{\mathbf{k}} \alpha_\sigma c^\dagger_{\mathbf{k},lp\sigma} + V^l_{\mathbf{k+q}} \beta_\sigma c^\dagger_{\mathbf{k+q},lp\sigma})/(E_{\mathbf{k}} - \varepsilon_p)]|0\rangle \qquad 7$$

where $C$ is the normalization constant, $\alpha_\uparrow/\alpha_\downarrow = i\sqrt{S_0^+/S_0^-} = ie^{i\eta}$, i. e. $\eta = \tan^{-1} S_0^y/S_0^x$, and

$$\frac{\beta_{\uparrow(\downarrow)}}{\alpha_{\uparrow(\downarrow)}} \approx \pm e^{i\xi}\left[1 + \frac{\sum_l (|V^l_{\mathbf{k}}|^2 - |V^l_{\mathbf{k+q}}|^2)}{|\Sigma| U (E - \varepsilon_p)}\right] \qquad 8$$

with $\xi = \arg(\Sigma)$. According to our calculation the second term on the right hand side of eq. (8) does not contribute to net dipole moment of the system. It does, however, give rise to dipole moment density waves (similar to charge density wave) of wave vector $\mathbf{q}$. Since it is not the focus of this paper, we will drop it hereafter and use the relation

$$\frac{\beta_{\uparrow(\downarrow)}}{\alpha_{\uparrow(\downarrow)}} \approx \pm e^{i\xi} \qquad 9$$

If $\gamma = 0$ or the two magnetic orders are in phase and can be treated as one, then all the bands are two-fold degenerate. This results in cancellation of electric dipole moment due to the difference in the relative sign between $\alpha_\sigma$ and $\beta_\sigma$ in two degenerate bands. In other words, if there is only one magnetic order ($\gamma = 0$) then there is no ferroelectricity. The following calculation is carried out under the assumption that $\gamma \neq 0$.

Anticipating the need of calculating the electric dipole moment per unit cell, we transform the wave function in eq. (7) to a localized wave function

$$w_j(\mathbf{r}) = C \sum_\sigma (\alpha_\sigma + \beta_\sigma e^{i\mathbf{q}\cdot\mathbf{R}_j})\{\psi_{d\sigma}(\mathbf{r} - \mathbf{R}_j) + \sum_l \frac{V_l}{E_0 - \varepsilon_p}[\psi_{pl\sigma}(\mathbf{r} - \mathbf{R}_j - \mathbf{r}_l) - \psi_{pl\sigma}(\mathbf{r} - \mathbf{R}_j + \mathbf{r}_l)]\}$$

$$- C \sum_{l,m,\sigma} \frac{V_l^3 \cos(\mathbf{q}\cdot\mathbf{r}_m)}{2(E_0 - \varepsilon_p)^3}\{(\alpha_\sigma e^{i\mathbf{q}\cdot\mathbf{r}_m} + \beta_\sigma e^{i\mathbf{q}\cdot(\mathbf{R}-\mathbf{r}_m)})[\psi_{pl\sigma}(\mathbf{r} - \mathbf{R}_j + 2\mathbf{r}_m - \mathbf{r}_l) - \psi_{pl\sigma}(\mathbf{r} - \mathbf{R}_j + 2\mathbf{r}_m + \mathbf{r}_l)]$$

$$+ (\alpha_\sigma e^{-i\mathbf{q}\cdot\mathbf{r}_m} + \beta_\sigma e^{i\mathbf{q}\cdot(\mathbf{R}+\mathbf{r}_m)})[\psi_{pl\sigma}(\mathbf{r} - \mathbf{R}_j - 2\mathbf{r}_m - \mathbf{r}_l) - \psi_{pl\sigma}(\mathbf{r} - \mathbf{R}_j - 2\mathbf{r}_m + \mathbf{r}_l)]\} \qquad 10$$

where subscript $\sigma$ is the spin index, $\psi_{pl\sigma}(\mathbf{r})$ is the wave function of the $p_{x(y)}$ orbital of the oxygen atoms and $\psi_d(\mathbf{r}) = A\langle \mathbf{r}|x^2 - y^2\rangle + B\langle \mathbf{r}|3z^2 - r^2\rangle$. Here we have made the approximation $1/(E_{\mathbf{k}} - \varepsilon_p) \approx 1/(E_0 - \varepsilon_p) - \sum_m [V_m/(E_0 - \varepsilon_p)]^2 \cos(\mathbf{q}\cdot\mathbf{r}_m) \cos[(2\mathbf{k} + \mathbf{q})\cdot\mathbf{r}_m]$ where $V_l = (\Lambda_l A + B/\sqrt{3})(pd\sigma)$, $E_0 = \varepsilon_d - J|\Sigma|/2 + (A^2 + B^2/3)(pd\sigma)^2/(\varepsilon_d - J|\Sigma|/2 - \varepsilon_p)$ in eq. (6).

The other interaction we considered was the spin-orbit coupling

$$H_1 = \lambda \mathbf{l}\cdot\mathbf{s} \qquad 11$$

Its effect can be seen in the following equations

$$\mathbf{l}\cdot\mathbf{s}\,|x^2 - y^2, \uparrow\rangle = i|xy, \uparrow\rangle + \frac{1}{2}(|zx, \downarrow\rangle - i|yz, \downarrow\rangle)$$

$$\mathbf{l}\cdot\mathbf{s}\,|x^2 - y^2, \downarrow\rangle = -i|xy, \downarrow\rangle - \frac{1}{2}(|zx, \uparrow\rangle + i|yz, \uparrow\rangle)$$

$$\mathbf{l}\cdot\mathbf{s}\,|3z^2 - r^2, \uparrow\rangle = -\frac{\sqrt{3}}{2}(|zx, \downarrow\rangle + i|yz, \downarrow\rangle)$$

$$\mathbf{l}\cdot\mathbf{s}\,|3z^2 - r^2, \downarrow\rangle = \frac{\sqrt{3}}{2}(|zx, \uparrow\rangle - i|yz, \uparrow\rangle). \qquad 12$$

The appearance of $t_{2g}$ orbitals demanded a little care. We defined the spin states parallel and antiparallel to $\mathbf{S}_j$ as

$$|P\rangle_j = \cos(\theta_j/2)|\uparrow\rangle_j + e^{i\phi_j}\sin(\theta_j/2)|\downarrow\rangle_j$$
$$|A\rangle_j = -e^{-i\phi_j}\sin(\theta_j/2)|\uparrow\rangle_j + \cos(\theta_j/2)|\downarrow\rangle_j \qquad 13$$

where $\theta_j$ and $\phi_j$ are the angles of $\mathbf{S}_j$. Since the states $|yz,P\rangle, |zx,P\rangle$ and $|xy,P\rangle$ are occupied by the $t_{2g}$ electrons, we should project the spin states in eq. (12) into antiparallel states. As a result, the spin-orbit coupling, when treated as a perturbation, modifies the $e_g$ orbitals in the following manner:

$$|x^2-y^2,\uparrow\rangle_j \to |x^2-y^2,\uparrow\rangle_j + \frac{\lambda}{\Delta_{cf}-J(3+|\Sigma|)/2}[-ie^{i\phi_j}\sin\frac{\theta_j}{2}|xy\rangle_j + \frac{1}{2}\cos\frac{\theta_j}{2}(|zx\rangle_j - i|yz\rangle_j)]|A\rangle_j$$

$$|x^2-y^2,\downarrow\rangle_j \to |x^2-y^2,\downarrow\rangle_j + \frac{\lambda}{\Delta_{cf}-J(3+|\Sigma|)/2}[-i\cos\frac{\theta_j}{2}|xy\rangle_j + \frac{1}{2}e^{i\phi_j}\sin\frac{\theta_j}{2}(|zx\rangle_j + i|yz\rangle_j)]|A\rangle_j$$

$$|3z^2-r^2,\uparrow\rangle_j \to |3z^2-r^2,\uparrow\rangle_j - \frac{\lambda}{\Delta_{cf}-J(3+|\Sigma|)/2}\frac{\sqrt{3}}{2}\cos\frac{\theta_j}{2}(|zx\rangle_j + i|yz\rangle_j)|A\rangle_j$$

$$|3z^2-r^2,\uparrow\rangle_j \to |3z^2-r^2,\downarrow\rangle_j + \frac{\lambda}{\Delta_{cf}-J(3+|\Sigma|)/2}\frac{\sqrt{3}}{2}e^{i\phi_j}\sin\frac{\theta_j}{2}(-|zx\rangle_j + i|yz\rangle_j)|A\rangle_j \qquad 14$$

where $\Delta_{cf}$ is the crystal field splitting between the $t_{2g}$ and $e_g$ states. We inserted the forms in eqs. (14) into eq. (8).

Now we are in a position to derive the electric polarization. The electric dipole moment of the j-th unit cell $\mathbf{P}_i$ is equal to $\int w_j^*(\mathbf{r})\mathbf{r}w_j(\mathbf{r})d^3\mathbf{r}$. Due to the symmetry of the atomic wave functions, the integrals that have to be considered are of the form ,
$\int \psi_d^*(\mathbf{r}-\mathbf{R}_j)\mathbf{r}\psi_{pl}(\mathbf{r}-\mathbf{R}_j \pm \mathbf{r}_l)d^3\mathbf{r}$. It turns out that due to the symmetry of the lattice, all the terms with $\int \psi_{x^2-y^2}^*(\mathbf{r}-\mathbf{R}_j)\mathbf{r}\psi_{pl}(\mathbf{r}-\mathbf{R}_j \pm \mathbf{r}_l)d^3\mathbf{r}$ cancel. What left are the terms with $\int \psi_{xy}^*(\mathbf{r}-\mathbf{R}_j)\mathbf{r}\psi_{pl}(\mathbf{r}-\mathbf{R}_j \pm \mathbf{r}_l)d^3\mathbf{r}$, i. e. those states generated the by spin-orbit interaction with the p-orbitals of oxygen atoms. This is compatible with the concept of bond-centered ferroelectricity in ref. 9. Here we define $\rho_{n,l} = \int \psi_n^*(\mathbf{r})\mathbf{r}\psi_{pl}(\mathbf{r}\pm \mathbf{r}_l)d^3\mathbf{r}$ where $n = yz, zx,$ or $xy$ and our result is the following:

$$\mathbf{P}_j = -2e\sum_l \delta_l|C\alpha_\uparrow|^2 \sin(2\mathbf{q}\cdot\mathbf{r}_l)\{\sin(\mathbf{q}\cdot\mathbf{R}_j+\xi)[(A-B)\sin\theta_j\cos\phi_j\boldsymbol{\rho}_{zx,l} - (A+B)\sin\theta_j\sin\phi_j\boldsymbol{\rho}_{yz,l}]$$
$$+ \cos(\mathbf{q}\cdot\mathbf{R}_j+\xi)[2A(\cos\eta\sin\theta_j\sin\phi_j - \sin\eta\sin\theta_j\cos\phi_j)\boldsymbol{\rho}_{xy,l}$$
$$+ (A-B)\cos\eta\cos\theta_j\boldsymbol{\rho}_{zx,l} - (A+B)\sin\eta\cos\theta_j\boldsymbol{\rho}_{yz,l}]\} \qquad 15$$

where $\delta_l = \lambda V_l^3/(E_0-\varepsilon_p)^3[\Delta_{cf}-J(3+|\Sigma|)/2]$. Note that $\boldsymbol{\rho}_{yz,x} = \boldsymbol{\rho}_{zx,y} = 0$, $\boldsymbol{\rho}_{yz,y} = \boldsymbol{\rho}_{zx,x} = \rho\hat{z}$, and $\boldsymbol{\rho}_{xy,x} = \rho\hat{y}$, $\boldsymbol{\rho}_{xy,y} = \rho\hat{x}$ with $\rho = \int \psi_{zx(y)}^*(\mathbf{r})z\psi_{px(y)}(\mathbf{r}\pm\mathbf{r}_{x(y)})d^3\mathbf{r}$. So the dipole moment is in the z-direction. Furthermore, one can relate the terms in eq. (15) to the $t_{2g}$ spins. For eample, $S_j^x = S_0^x \sin\theta_j\cos\phi_j$, $S_j^y = S_0^y \sin\theta_j\sin\phi_j$ and $S_j^z = S_0^z \cos\theta_j$. Let us give an estimation of the magnitude of the resulting polarization. $\Delta_{cf} \approx 3eV$, $J \approx 2eV$, $\varepsilon_d - 3J/2 - \varepsilon_p \approx 2eV$, $(pd\sigma)t \approx 0.5eV$, $\lambda \approx 0.05eV$. Since the volume of a unit cell is approximately $100\text{Å}^3/2$, $|C\alpha_\uparrow|^2 \approx 1/4$ and $\rho \approx 1\text{Å}$, $p_z$ is of the order $100\mu C/m^2$ which is compatible with experimental data.

In order to see more clearly the meaning of eq. (15), we considered a very simple case. Let $\gamma = \pi/2$, $\eta = 0$ and $A = 1, B = 0$ or $A = 0, B = 1$. These meant that the phase difference between the magnetic orders of $S^z$ and $S^{x(y)}$ is $\pi/2$, $S_0^y = 0$, and the magnitude of the total spin is a constant. Furthermore, either $|x^2-y^2\rangle$ or $|3z^2-r^2\rangle$ is considered. Then eq. (15) is simplified to

$$\mathbf{P}_j = \frac{4e\rho|C\alpha_\uparrow|^2\lambda(pd\sigma)^3}{S_0^z S_0^x (E_0-\varepsilon_p)^3[\Delta_{cf}-J(3+|\Sigma|)/2]}[(S_{j+x}^z - S_{j-x}^z)S_j^x - (S_{j+x}^x - S_{j-x}^x)S_j^z]\hat{z} \qquad 16$$

where $S_{j+x(y)}$ is the spin at $R_j \pm 2r_{x(y)}$. We can rewrite eq. (16) to give a compact expression of the electric polarization $\boldsymbol{p} \sim \hat{z}[M_z \nabla \cdot \boldsymbol{M} - (\boldsymbol{M} \cdot \nabla) M_z]$. The phenomenological model[14] gives a similar form: $\boldsymbol{p} \sim [\boldsymbol{M}(\nabla \cdot \boldsymbol{M}) - (\boldsymbol{M} \cdot \nabla)\boldsymbol{M}]$. On the one hand, what we have calculated is simpler. Only the dynamics on the *ab* plane was considered and this is the reason why the polarization is pointing in the z-direction. We must add that this approximation is compatible with the structures of the systems such as $TbMnO_3$. On the other hand, there is significant complication in our calculation, namely the dynamics and anisotropy of orbitals. It is remarkable that we reached the expression of ref. 14. There had been a long history[1] of studying the ferroelectricity phase transition with Landau's theory. If Mostovoy's model[14], which is supported by our microscopic calculation, is correct, then there should be a term of the form $g\boldsymbol{p} \cdot [\boldsymbol{M}(\nabla \cdot \boldsymbol{M}) - (\boldsymbol{M} \cdot \nabla)\boldsymbol{M}]$ in the thermodynamic potential where $g$ is a constant. This implies that the phase transition is an improper ferroelectric phase transition[15,16] and the magnetic orders, instead of the electric polarization order should be the order parameters. All the thermodynamic properties should be understood thusly.

As for the microscopic mechanism, the spin-current calculation proposed by Katsura[10] et. al. gave a electric dipole moment of the form $\hat{e} \times (S_1 \times S_2)$ with $\hat{e}$ being the direction of M-O-M bond. This is equivalent to our result in eqs. (15) and (16) and Mostovoy's form[14] if the expressions in eqs. (2) are used. If there is only one sinusoidal magnetic order, then there is no spin current and our calculation showed that there is no electric polarization. Even though Katsura et. al.[10] considered $t_{2g}$ orbitals and their $\pi$-bonds with oxygen atoms and two noncollinear spins, while we $e_g$ orbitals, $\sigma$-bonds and magnetic orders, the spin-current interpretation certainly accord with our calculation.

In conclusion, we have shown that there is indeed a direct relationship between magnetic orders and ferroelectricity with the spin-orbit interaction being the link and the anisotropy of orbitals as the hotbed of dipole moments.

The author is in debt to Chung-Yu Mou and Di-Jing Huang for inspiring discussion and benefit from the activities of SCES and "spin-related physics in condensed matter" focus groups of NCTS, Taiwan. This work is supported in part by the National Science Council under the contract NSC 94-2112-M-002-46.